\documentclass[11pt]{article}
\usepackage{graphicx}
\usepackage{amssymb}
\usepackage{amscd}
\usepackage{mathrsfs}
\usepackage{longtable,lscape}
\usepackage{amsthm}
\usepackage{amsfonts}
\usepackage{amsmath}
\usepackage{bbm}
\usepackage{float}
\usepackage{url}
\usepackage{hyperref}

\newcommand{\captionfonts}{\footnotesize}
\makeatletter  
\long\def\@makecaption#1#2{%
  \vskip\abovecaptionskip
  \sbox\@tempboxa{{\captionfonts #1: #2}}%
  \ifdim \wd\@tempboxa >\hsize
    {\captionfonts #1: #2\par}
  \else
    \hbox to\hsize{\hfil\box\@tempboxa\hfil}%
  \fi
  \vskip\belowcaptionskip}
\makeatother 
\begin{document}
\title{Entanglement in Cognition violating Bell Inequalities \\ Beyond Cirel'son's Bound}
\author{Diederik Aerts, Jonito Aerts Argu\"{e}lles,
Lester Beltran \\ Suzette Geriente\footnote{Center Leo Apostel for Interdisciplinary Studies, 
        Free University of Brussels (VUB), Krijgskundestraat 33,
         1160 Brussels, Belgium; email addresses: diraerts@vub.ac.be,lestercc21@gmail.com,sgeriente83@yahoo.com} 
        $\,$ and $\,$  Sandro Sozzo\footnote{School of Business and Centre IQSCS, University of Leicester, University Road, LE17RH Leicester, United Kingdom; email address: ss831@le.ac.uk}              }
\date{}
\maketitle
\begin{abstract}
\noindent
We present the results of two tests where a sample of human participants were asked to make judgements about the conceptual combinations {\it The Animal Acts} and {\it The Animal eats the Food}. Both tests significantly violate the Clauser-Horne-Shimony-Holt version of Bell inequalities (`CHSH inequality'), thus exhibiting manifestly non-classical behaviour due to the meaning connection between the individual concepts that are combined. We then apply a quantum-theoretic framework which we developed for any Bell-type situation and represent empirical data in complex Hilbert space. We show that the observed violations of the CHSH inequality can be explained as a consequence of a strong form of `quantum entanglement' between the component conceptual entities in which both the state and measurements are entangled. We finally observe that a quantum model in Hilbert space can be elaborated in these Bell-type situations even when the CHSH violation exceeds the known `Cirel'son bound', in contrast to a widespread belief. These findings confirm and strengthen the results we recently obtained in a variety of cognitive tests and document and image retrieval operations on the same conceptual combinations.  
\end{abstract}

\section{Introduction\label{intro}}
In physics, the violation of the so-called `Bell inequalities' is generally maintained to prove that quantum entities exhibit specific aspects, as  `contextuality', `entanglement' and `nonseparability', which cannot be explained in terms of the mathematical structures that are typically used in classical physics. This contextuality also persists for far away quantum entities, an aspect known as `nonlocality' after Bell's seminal paper \cite{bell1964,clauser1969}. The consistent violation of Bell inequalities in physics tests performed on a variety of quantum entities, in addition to confirming the predictions of quantum theory, in particular reveals that the statistics of repeated experiments shows connections between quantum entities that cannot be represented within a classical probabilistic model satisfying the axioms of Kolmogorov (`Kolmogorovian probability') (see, e.g., \cite{pitowsky1989}).

The setting of a Bell-type test is relatively simple, if formulated in the so-called `EPR-Bohm form' \cite{epr1935,bohm1951}. We briefly review it here, as Bell-type tests are relevant to the purposes of the present paper. One considers a composite physical entity ${\mathscr S}_{12}$, prepared in an initial state $p$ and such that two individual entities ${\mathscr S}_1$ and ${\mathscr S}_2$ that have interacted in the past but are now far away (`space-like separation') can be recognized as parts of ${\mathscr S}_{12}$. Then, 4 coincidence experiments $AB$, $AB'$, $A'B$ and $A'B'$ are performed on ${\mathscr S}_{12}$ which consist in performing measurements $e_{A}$, with outcomes $A_1$ and $A_2$, and $e_{A'}$ with outcomes $A'_1$ and $A'_2$, on entity ${\mathscr S}_1$, and measurements $e_{B}$, with outcomes $B_1$ and $B_2$, and $e_{B'}$ with outcomes $B'_1$ and $B'_2$, on entity ${\mathscr S}_2$. If the measurement outcomes can be only $- 1$ or $+1$, then the expected values of $AB$, $AB'$, $A'B$ and $A'B'$ become the correlation functions $E(A,B)$, $E(A,B')$, $E(A',B)$ and $E(A',B')$, respectively, and a specific inequality, which is the `Clauser-Horne-Shimony-Holt version of Bell inequalities' (briefly, `CHSH inequality') 
\begin{equation} \label{chsh}
-2 \le E(A',B')+E(A',B)+E(A,B')-E(A,B) \le 2
\end{equation}
can be formulated \cite{clauser1969}. John Bell put forward the type of inequality carrying his name \cite{bell1964}, with the aim of making experimental tests possible with respect to the strange quantum phenomenon revealed in the Einstein Podolsky Rosen paradox reasoning \cite{epr1935} called `entanglement', i.e. if this phenomenon would not be present the inequality would not be violated. The violation of Bell's inequality can also be connected with the non-existence of a Kolmogorovian probability model for the joint entity \cite{pitowsky1989}, but in the investigation we put forward here our interest goes to the original situation for which the inequalities were formulated, i.e. the connections possible between individual entities recognizable within a joint entity. It is indeed the case that quantum theory makes specific predictions about Equation (\ref{chsh}). More specifically, one can choose the individual entities, e.g., pair of spin 1/2 quantum particles, states, e.g., entangled state (typically, the singlet spin state), and observables, e.g., pairs of spin along different directions, in such a way that Equation (\ref{chsh}) is violated \cite{bell1964,clauser1969}. Moreover, one can show that the `CHSH factor' $\Delta_{CHSH}=E(A',B')+E(A',B)+E(A,B')-E(A,B)$ is bound in quantum theory by the numerical value $\Delta_{QMC}=2\sqrt{2} \approx 2.83$, known as the `Cirel'son bound'  \cite{cirelson1980,cirelson1993}.  It is important to observe, at this stage, that the existence of such a `quantum bound' is not trivial because, from a mathematical point of view, the CHSH factor can take any value between $-4$ and $+4$. In addition, the derivation of the Cirel'son bound explicitly refers to measurements $e_{XY}$, corresponding to coincidence experiments $XY$, $X=A,A'$, $Y=B,B'$, on the composite entity that are represented by product self-adjoint operators whenever the states of the composite entity ${\mathscr S}_{12}$ are represented by the unit vectors of the tensor product Hilbert space of two Hilbert spaces of which their unit vectors represent the states of the possible to be recognized entities ${\mathscr S}_1$ and ${\mathscr S}_2$. This remark will play an important role in what follows.

In the last two decades, several theoretical studies, mainly inspired by quantum computation and quantum information and their flourishing applications, have deeply analysed and extended Bell inequalities (see, e.g., \cite{genovese2005,brunner2014}). In addition, numerous empirical tests have followed the seminal experiments of Aspect and his collaborators \cite{aspect1982a,aspect1982b}. All empirical tests
that have been performed so far strongly confirm the predictions of quantum theory (see, e.g., \cite{tittel1998,weihs1998,vienna2013,urbana2013}). The natural consequence of these results is that  `entanglement', i.e. unavoidable connection  between the individual entities recognizable in a composite quantum entity, give rise to empirically observable phenomena, so much that, in particular,  entanglement  is nowadays considered as one of the fingerprints of quantum theory.

Growing theoretical and empirical research reveals that  several quantum structures, such as `contextuality', `entanglement', `indistinguishability', `interference' and `superposition', also manifest themselves in other domains than micro-physics, which include, in particular, cognition (e.g., human probability and similarity judgements, decision-making and visual perception), socio-economic domains (economics and finance) and information systems (computer science and artificial intelligence) (see, e.g., \cite{aerts2009,pothosbusemeyer2009,khrennikov2010,busemeyerbruza2012,aertsbroekaertgaborasozzo2013,aertsgaborasozzo2013,
havenkhrennikov2013,kwampleskacbusemeyer2015,dallachiaragiuntininegri2015a,dallachiaragiuntininegri2015b,melucci2015,
aertsetal2018e,haven2018,arguelles2018,arguellessozzo2019} and references therein). In particular, interesting results have been obtained in the identification of quantum structures and, in particular, quantum entanglement, in the combination of natural concepts and applied computer science domains, as information retrieval and natural language processing (see, e.g., \cite{schmittetal2008,aerts2010,bruza2009,bruzakittonelsonmcevoy2009,coecke2010,piwowarskietal2010,frommholzetal2010,
aertssozzo2011,ingo2011,bucciomeluccisong2011,aertssozzo2014a,aertssozzo2014b,bruzakittorammsitbon2015,aertsIJTP,
gronchistrambini2017,beltrangeriente2018,
aertsbeltrangerientesozzo2019} and references therein). In this regard, our research team has produced both theoretical and empirical research on how to identify entanglement in conceptual combinations and document retrieval applications. 

At an empirical level, in particular, we performed both cognitive tests on human participants \cite{aertssozzo2011}, document retrieval tests on structured corpuses of documents 
 \cite{beltrangeriente2018} and image retrieval tests on the web \cite{arguelles2018}. In these tests, we considered the composite conceptual entity {\it The Animal Acts} as a combination of the individual conceptual entities {\it Animal} and {\it Acts}. We performed an additional document retrieval test on the web in which we considered the  composite conceptual entity {\it The Animal eats the Food} as a combination of the individual conceptual entities {\it Animal} and {\it Food} \cite{aerts2010}. In all these tests, we observed a systematic violation of the CHSH inequality, accompanied by a systematic violation of a property  called the `marginal law'. 
 
 In particular, the violation of the marginal law, which is believed not to occur in principle in Bell-type tests in quantum physics -- we come back to this later, led  us to study in detail a theoretical problem which is usually overlooked in physics, the `identification  problem', i.e. the problem of recognising individual entities of a composite entity by performing on the latter measurements that resemble the typical coincidence experiments of Bell-type tests \cite{aertssozzo2014a,aertssozzo2014b}. We constructed a general theoretical framework to model any Bell-type situation in the Hilbert space formalism of quantum theory \cite{aertsetal2019}.
In this framework, one explicitly applies the prescription of the standard formalism of quantum theory that the joint entity needs to be modelled in a complex Hilbert space of which the dimension is given by the number of outcomes of the defining measurements, which is hence $\mathbb{C}^{4}$. Only then as a secondary step, with respect to the attempt to `recognize' 
entities in the joint entity, one considers possible isomorphisms with another Hilbert space, namely, the one built as the tensor product $\mathbb{C}^{2} \otimes \mathbb{C}^{2}$ of two-dimensional complex Hilbert spaces which need to describe the two possible 
entities, since the defining measurements to recognize these 
entities have two outcomes. Like we analyzed in detail in \cite{aertssozzo2014a,aertssozzo2014b}, there is no unique isomorphism between $\mathbb{C}^{4}$ and $\mathbb{C}^{2} \otimes \mathbb{C}^{2}$ and this is the reason that from a mathematical point of view there are different ways to account for entanglement being present within the joint entity with respect to the 
entities
to be recognized individual entities. Essentially, entanglement manifests on the level of the probabilities of a joint measurements not being able to be written as products of probabilities on the level of the component measurements of this joint measurement and hence it is a property of the relation between the joint measurements and their components. Only when an extra symmetry is present connecting different joint measurements the entanglement of these different joint measurements can be captured in a state of the joint entity. This symmetry can be detected by verifying whether the marginal law corresponding to the considered joint measurements is satisfied, in that case, the entanglement of these joint measurements can together be captured in a state of the joint entity. If this is true for all joint measurements one can prove that there is only one isomorphism connecting $\mathbb{C}^{4}$ with $\mathbb{C}^{2} \otimes \mathbb{C}^{2}$ and in that case $\mathbb{C}^{4}$ can be substituted by $\mathbb{C}^{4}$ with $\mathbb{C}^{2} \otimes \mathbb{C}^{2}$ and the joint entity can be directly modeled in $\mathbb{C}^{2} \otimes \mathbb{C}^{2}$. With respect to `entanglement', a property of the structure of the probabilities of a joint measurement with respect to the probabilities of the component measurements, this situation is exceptional and definitely not the general one. In \cite{aertssozzo2014b} we work out in detail several experimental examples and how how entanglement can generally only be located in the joint measurements and only exceptionally in the state of the joint entity. It is widely accepted that in the typically considered quantum physics situations where entanglement is being measured and studied this exceptional symmetry is always present and hence apriori the tensor product Hilbert space is taken to be the Hilbert space to model the joint entity, although already the first experiments of Alain Aspect \cite{aspect1982a} showed violations of the marginal law, confirmed manifestly by later experiments \cite{weihs1998}. For the cognition experiments that we consider in the present article where joint measurements are performed on joint concepts {\it The Animal Acts} and {\it The Animal eats the Food} the symmetry is not fulfilled and hence no unique isomorphism exists between $\mathbb{C}^{4}$ and $\mathbb{C}^{2} \otimes \mathbb{C}^{2}$. The new quantum-theoretic perspective enabled modelling of both cognitive and web tests on {\it The Animal Acts} and introduced a novel mathematical ingredient in the modelling  of Bell-type situations, namely, `entangled measurements' \cite{arguellessozzo2019,aertssozzo2014a,aertssozzo2014b,aertsbeltrangerientesozzo2019}. 

Entangled measurements have recently been applied with success to perform quantum computation and information tasks (see, e.g., \cite{brunner2014}). We show in the present paper that `entanglement measurement also allow to identify the presence of quantum entanglement also in Bell-type tests where the violation of the Cirel'son bound occurs in addition to the violation of the CHSH inequality'. More specifically, we present the results of two cognitive tests, one on {\it The Animal Acts} and the other on {\it The Animal eats the Food} situations, that we have recently performed on a sample of 81 participants. In both tests, a significant violation of the CHSH has been observed. Furthermore, while {\it The Animal Acts} test has revealed a result very close to 
the Cirel'son bound, {\it The Animal eats the Food} test has revealed a significant violation of the latter bound -- a result which we had already been obtained in the web tests above on {\it The Animal Acts} \cite{beltrangeriente2018,aertsbeltrangerientesozzo2019}. We successfully apply and refine the quantum-theoretic framework to {\it The Animal Acts} and {\it The Animal eats the Food} situations and get a confirmation that a `strong form of quantum entanglement' exists between the respective individual concepts which involves both states and measurements'. Moreover, we get a further confirmation that `such a form of entanglement captures the deeply non-classical meaning connections existing between those individual concepts'. Finally, we prove that `quantum entanglement also exists beyond Cirel'son bound, but it is a joint effect of state-entanglement and measurement-entanglement. This final result is relevant, in our opinion, as it confutes the widespread belief that one cannot model situations exceeding Cirel'son in the Hilbert space formalism in Hilbert space (see, e.g., \cite{cirelson1980,cirelson1993,brunner2014}). As such, this final result may also have an impact on the foundations of quantum theory.

For the sake of completeness, we summarise the content of this paper in the following.

In Section \ref{history}, we briefly review the theoretical and empirical results that our research team has obtained in the last years on the identification of quantum entanglement in conceptual combinations. In Section \ref{data}, we report and analyse the data collected in two cognitive tests that we have recently performed on human participants about the conceptual combinations  {\it The Animal Acts} and {\it The Animal eats the Food}. In Section \ref{model}, we explicate, refine and particularise to {\it The Animal Acts} and {\it The Animal eats the Food} situations the general quantum-theoretic framework that we have developed in \cite{aertsetal2019} to model any Bell-type situation. In Section \ref{quantumrep}, we show that the quantum-theoretic framework allows to faithfully represent empirical data in Section \ref{data}. Finally, in Section \ref{bound}, we explain how and why a violation of the CHSH inequality which also exceed the Cirel'son bound may indicate the presence of a stronger form of quantum entanglement involving both states and measurements.

\section{A brief history of Bell-type tests in conceptual combinations\label{history}}
We review in this section the results that we have obtained in the last decade on the identification of entanglement in the combination of natural concepts and related applied domains, as information retrieval and natural language processing. For a detailed description of the tests and a complete analysis of the obtained results, the reader is referred to the papers cited below (see also Section \ref{data}).

In the first empirical test of entanglement involving concepts, we studied the combination {\it The Animal eats the Food}, which we regarded as a combination of the individual concepts {\it Animal} and {\it Food} \cite{aerts2010}. We considered different items of {\it Animal}, namely, {\it Cat}, {\it Cow}, {\it Horse} and {\it Fish}, and different items of {\it Food}, namely, {\it Grass}, {\it Meat}, {\it Fish} and {\it Nuts}, and combined them to form all possible combinations, i.e. {\it The Cow eats the Grass}, {\it The Cat eats the Meat}, {\it The Cat eats the Fish}, {\it The Squirrel eats the Nuts}, and so on, which were considered as items of the combination {\it The Animal eats the Food}. Next, we split the 16 combinations obtained in this way into 4 groups of 4 combinations each in order to reproduce the 4 coincidence experiments described in Section \ref{intro}. For example, coincidence experiment $AB$ had 4 outcomes, {\it The Cat eats the Grass}, {\it The Cat eats the Meat}, {\it The Cow eats the Grass} and {\it The Cow eats the Meat}; the other experiments were constructed in an analogous way. We used the World Wide Web as a conceptual space and counted co-occurrence of words, e.g., ``cow'' and ``grass'', ``cat'' and ``fish'', and so on, in document retrieval operations by means of the `Google' search engine. We collected relative frequencies of co-occurrences which we considered, in the large number limit, as joint probabilities and inserted them into the correlation functions in Equation (\ref{chsh}). With our surprise, we found $\Delta_{CHSH}=2.86$, hence a significant violation of the CHSH inequality. We  put forward
the presence of some type of entanglement between the concepts {\it Animal} and {\it Food} was responsible of the observed deviation from the CHSH inequality. At that time, we also noticed without however deepening it, the simultaneous violation of Cirel'son bound.

This empirical finding was really exciting for us, also because it was partially unexpected, thus we decided to deepen the investigation on entanglement in conceptual combinations and performed a cognitive test in which the direct responses of a sample of participants were collected \cite{aertssozzo2011}. In this case, we analysed the conceptual combination {\it The Animal Acts}, which we regarded as a combination of the individual concepts {\it Animal} and {\it Acts} -- ``acts'' refers here to the action of an animal emitting a sound. We again considered different items of {\it Animal}, namely, {\it Horse}, {\it Bear}, {\it Tiger} and {\it Cat}, and different items of {\it Atcs}, namely, {\it Growls}, {\it Whinnies}, {\it Snorts} and {\it Meows}, and combined them to form all possible combinations, i.e. {\it The Horse Growls}, {\it The Bear Whinnies}, {\it The Tiger Snorts}, {\it The Cat Meows}, and so on, which were considered as items of the combination {\it The Animal Acts}. Then, we split the 16 combinations obtained in this way into 4 groups of 4 combinations each in order to reproduce the 4 coincidence experiments in Section \ref{intro}. For example, coincidence experiment $AB$ had 4 outcomes, {\it The Horse Growls}, {\it The Horse Whinnies}, {\it The Bear Growls}, {\it The Bear Whinnies}; the other experiments were constructed in a similar way. After reading an introductory text, a sample of 81 respondents had to fill in a questionnaire where, in each coincidence experiment, they had to choose which item in a list of 4 items they judged as a `good example' of the conceptual combination {\it The Animal Acts}. Relative frequencies of positive responses were considered,  in the large number limit, as joint probabilities and inserted into  the correlation functions in Equation (\ref{chsh}). Also in this case, we found a significant violation the CHSH inequality, $\Delta_{CHSH}=2.42$, which we again interpreted as the consequence of a form of entanglement in the combination {\it The Animal Acts}, due to the connections of meaning between the component concepts {\it Animal} and {\it Acts}.

The numerical value $2.42$ obtained in the cognitive test on {\it The Animal Acts} resembled the values found in Bell-type tests in quantum physics (see Section \ref{intro}). To make the analogy with physics more convincing, we then decided to construct an explicit quantum-theoretic model in Hilbert space of empirical data \cite{aertssozzo2014a}. But, we realised at once that we could not reproduce empirical data on {\it The Animal Acts} test by representing the initial state of {\it The Animal Acts} by an entangled state and the 4 measurements by product measurements, as one would be tempted to do in analogy with a quantum representation in physics. We discovered that the main reason of this impossibility was that, if one wants to model {\it The Animal Acts} situation starting from the modelling of {\it Animal} and {\it Acts}, hence using the usual tensor product representation, then the `marginal law of probability' has to be satisfied, as it occurs in quantum physics, whereas it was systematically violated in the cognitive test. This difference between physical and cognitive realms led us to study in detail a problem that is generally overlooked in physics, the `identification problem', that is, the problem of recognising individual entities, e.g., concepts, from measurements performed on composite entities, e.g., conceptual combinations. The investigation led us to develop a novel quantum-theoretic framework for any Bell-type situation, which allowed us to recognise systematic analogies and differences between violations of Bell inequalities in physics and cognition \cite{aertsetal2019}.\footnote{A closer analogy between Bell-type situations in physical and cognitive realms was identified in a cognitive test we performed on the conceptual combination {\it Two Different Wind Directions}, where the marginal law was not violated and we could represent data in Hilbert space using entangled states and product measurements \cite{aertsetal2018a,aertsetal2018b}.} But, the quantum-theoretic framework also introduced a new fundamental element in the modelling of such situations, namely, the empirically justified use of `entangled measurements' in addition to entangled states. Entangled measurements were indeed successfully used in the quantum representation of {\it The Animal Acts} situation \cite{aertssozzo2014a}.

In the meanwhile, we had also started systematically looking into possible applications of quantum structures in information retrieval and natural language processing, where representation of meaning entities, like concepts and conceptual combinations, play a crucial role (see, e.g., \cite{aertsIJTP,aertsetal2018e}). In particular, we considered again {\it The Animal Acts} situation in a document retrieval test in which we collected data on co-occurrence of words, as we had done for {\it The Animal eats the Food} situation, but using known corpuses of documents instead of web search engines -- corpuses of documents provide more reliable word counts than search engines  \cite{beltrangeriente2018}. We used  `Google Books', `Corpus of Contemporary American English (COCA)' and `News On Web (NOW)' as corpuses of documents and performed the coincidence experiments described in Section \ref{intro}, finding 3.41, 3.00 and 3.33, respectively in the factor appearing in Equation (\ref{chsh}). These empirical findings were statistically significant and, more important, they were completely unexpected. Indeed, not only we had obtained consistent violations of the CHSH inequality across all corpuses of documents, but the numerical value of the violation systematically exceeded the Cirel'son bound. According to a widespread belief, a violation of the CHSH inequality which also violates Cirel'son bound cannot be modelled in Hilbert space formalism of quantum theory, hence does not represent a `quantum effect' \cite{brunner2014}. On the other side, we noticed that the derivation of this bound does not consider entangled measurements, but only product measurements \cite{cirelson1980,cirelson1993}. If one allows entangled measurements, then it is in principle possible to model in Hilbert space any violation of the CHSH inequality whose numerical value lies in the mathematically permitted interval from $-4$ to $+4$. Indeed, we applied in \cite{aertsbeltrangerientesozzo2019} the general quantum-theoretic framework and faithfully represented the web data presented in \cite{beltrangeriente2018}.

Finally, in an independent but related investigation, we performed a web test on artefacts of visual perception, such as images, using `Google Images' as a web search engine to count images representing items of the conceptual combination {\it The Animal Acts} \cite{arguelles2018}. Also in this case, we obtained the value $\Delta_{CHSH}=2.41$ for the CHSH factor in Equation (\ref{chsh}), which again indicated the presence of entanglement in visual perception. The corresponding quantum-theoretic modelling enabled faithful representation of empirical data and confirmed that a strong form of entanglement, involving both states and measurements, exists between the component concepts {\it Animal} and {\it Acts}, again due to their meaning connections.

We now intend to complete the investigation above and consider two cognitive tests on the conceptual combinations  {\it The Animal Acts} and {\it The Animal eats the Food} with the aim of `identifying the presence of quantum entanglement beyond the limits imposed by Cirel'son bound'. We will dedicate the next sections to this purpose.

\section{New empirical results\label{data}}
We report in this section the details of the two cognitive tests that we have performed on the conceptual combinations {\it The Animal Acts} and {\it The Animal eats the Food}. As we will see, their results substantially confirm the empirical patterns highlighted in Section \ref{history}.

Let us start by {\it The Animal Acts} test. As anticipated in Section \ref{history}, we consider the concept {\it The Animal Acts} as a combination of the individual concepts {\it Animal} and {\it Acts}, where ``acts'' refers to the sound, or noise, produced by an animal. Next, we consider two pairs of items of {\it Animal}, namely, ({\it Horse}, {\it Bear}) and ({\it Tiger}, {\it Cat}), and two pairs of items of {\it Acts}, namely, ({\it Growls}, {\it Whinnies}) and ({\it Snorts}, {\it Meows}). We are now ready to illustrate the test.

A sample of 81 individuals were presented in a `within subjects design' a questionnaire which contained 4 coincidence experiments $AB$, $AB'$, $A'B$ and $A'B'$ whose setting was similar to the typical setting of a Bell-type test sketched in Section \ref{intro}. More specifically, participants were preliminarily asked to read an `introductory text' where the concepts under study were introduced and a description of the tasks involved in the judgement test was provided. Then, in each coincidence experiment participants were asked  to choose which item in a list of 4 items they judged as a good example of the conceptual combination {\it The Animal Acts}.

In coincidence experiment $AB$, participants had to choose among the 4 items: 

($A_1B_1$) {\it The Horse Growls}

($A_2B_2$) {\it The Bear Whinnies}

($A_1B_2$) {\it The Horse Whinnies}

($A_2B_1$) {\it The Bear Growls}

If the response was ($A_1B_1$) or ($A_2B_2$), then experiment $AB$ was attributed outcome $+1$; if the response was ($A_1B_2$) or ($A_2B_1$), then experiment $AB$ was attributed outcome $-1$. 

In coincidence experiment $AB'$, participants had to choose among the 4 items: 

($A_1B'_1$) {\it The Horse Snorts}

($A_1B'_2$) {\it The Horse Meows}

($A_2B'_1$) {\it The Bear Snorts}

($A_2B'_2$) {\it The Bear Meows}

If the response was ($A_1B'_1$) or ($A_2B'_1$), then experiment $AB'$ was attributed outcome $+1$; if the response was ($A_1B'_2$) or ($A_2B'_1$), then experiment $AB'$ was attributed outcome $-1$. 

In coincidence experiment $A'B$, participants had to choose among the 4 items: 

($A'_1B_1$) {\it The Tiger Growls}

($A'_1B_2$) {\it The Tiger Whinnies}

($A'_2B_1$) {\it The Cat Growls}

($A'_2B_2$) {\it The Cat Whinnies}

If the response was ($A'_1B_1$) or ($A'_2B_2$), then experiment $A'B$ was attributed outcome $+1$; if the response was ($A'_1B_2$) or ($A'_2B_1$), then experiment $A'B$ was attributed outcome $-1$. 

Finally, in coincidence experiment $A'B'$, participants had to choose among the 4 items: 

($A'_1B'_1$) {\it The Tiger Snorts}
 
($A'_1B'_2$) {\it The Tiger Meows}

($A'_2B'_1$) {\it The Cat Snorts}

($A'_2B'_2$) {\it The Cat Meows}

If the response was ($A'_1B'_1$) or ($A'_2B'_2$), then experiment $A'B'$ was attributed outcome $+1$; if the response was ($A'_1B'_2$) or ($A'_2B'_1$), then experiment $A'B'$ was attributed outcome $-1$. 

For each coincidence experiment $AB$, $AB'$, $A'B$ and $A'B'$, we collected the relative frequencies of the obtained responses which we considered, in the large number limit, as the probability $\mu(A_iB_j)$,  $\mu(A_iB'_j)$, $\mu(A'_iB_j)$ and $\mu(A'_iB_j)$ that the outcome $A_iB_j$, $A_iB'_j$, $A'_iB_j$ and $A'_iB'_j$, $i,j=1,2$, respectively, is obtained in the corresponding experiment. Table \ref{tab1} reports the judgement probabilities computed in this way. Referring to these probabilities, we can then calculate the expectation values, or correlation functions, of coincidence experiments $AB$, $AB'$, $A'B$ and $A'B'$,  as follows:
\begin{eqnarray}
E(AB)&=&\mu(A_1B_1)-\mu(A_1B_2)-\mu(A_2B_1)+\mu(A_2B_2)=-0.8025 \label{AB_Test_1} \\
E(AB')&=&\mu(A_1B'_1)-\mu(A_1B'_2)-\mu(A_2B'_1)+\mu(A_2B'_2)=0.4568 \label{AB'_Test_1}  \\
E(A'B)&=&\mu(A'_1B_1)-\mu(A'_1B_2)-\mu(A'_2B_1)+\mu(A'_2B_2)=0.7037 \label{A'B_Test_1} \\
E(A'B')&=&\mu(A'_1B'_1)-\mu(A'_1B'_2)-\mu(A'_2B'_1)+\mu(A'_2B'_2)=0.8765 \label{A'B'_Test_1}
\end{eqnarray}
Inserting Equations (\ref{AB_Test_1})--(\ref{A'B'_Test_1}) into Equation (\ref{chsh}), we get
\begin{equation}
\Delta_{CHSH}=E(A',B')+E(A',B)+E(A,B')-E(A,B)=2.7901
\end{equation}
\begin{table} \label{tab1}
\centering
\begin{tabular}{|c |c | c | c| c| }
\hline
Experiment $AB$ & \emph{Horse Growls} & \emph{Horse Whinnies} & \emph{Bear Growls} & \emph{Bear Whinnies}\\
Probability & $\mu(A_1B_1)=0.0494$ & $\mu(A_1B_2)=0.1235$ & $\mu(A_2B_1)=0.7778$ & $\mu(A_2B_2)=0.0494$  \\
\hline
 Experiment $AB'$  & \emph{Horse Snorts} & \emph{Horse Meows} & \emph{Bear Snorts} & \emph{Bear Meows}\\
Probability & $\mu(A_1B'_1)=0.7160$ & $\mu(A_1B'_2)=0.0494$ & $\mu(A_2B'_1)=0.2222$   & $\mu(A_2B'_2)=0.0123$ \\
\hline
Experiment $A'B$  &  \emph{Tiger Growls} & \emph{Tiger Whinnies} & \emph{Cat Growls} & \emph{Cat Whinnies}\\
Probability &   $\mu(A'_1B_1)=0.7778$ & $\mu(A'_1B_2)=0.0864$ & $\mu(A'_2B_1)=0.0617$  &  $\mu(A'_2B_2)=0.0741$ \\
\hline
Experiment $A'B'$  &   \emph{Tiger Snorts} & \emph{Tiger Meows} & \emph{Cat Snorts} & \emph{Cat Meows}\\
Probability &  $\mu(A'_1B'_1)=0.0864$ & $\mu(A'_1B'_2)=0.0617$ & $\mu(A'_2B'_1)=0.0247$ & $\mu(A'_2B'_2)=0.8272$\\
\hline
\end{tabular}
\caption{The data collected in coincidence experiments on entanglement in the conceptual combination {\it The Animal Acts}.}
\end{table}
The numerical value $2.7901\approx 2.79$ exceeds the classical limit imposed by the CHSH inequality and is slightly below the Cirel'son bound. However, while the deviation from the value $2$ is statistically significant (p-value $1.44*10^{-5}$), the deviation from the value $2\sqrt{2}$ is not significant (p-value $0.4151$), which entails that {\it The Animal Acts} data show a 
non definite behaviour with respect to the Cirel'son bound. This empirical pattern confirms and strengthens the results obtained in \cite{aertssozzo2011} where a violation of the CHSH inequality was observed. This non-classical behaviour admits the presence of entanglement between the concepts {\it Animal} and {\it Acts} as a natural explanation, because {\it Animal} and {\it Acts} are connected by meaning and this connection gives rise to statistical correlations, those expressed by Equations (\ref{AB_Test_1})--(\ref{A'B'_Test_1}).

Let us now come to {\it The Animal eats the Food} test. As mentioned in Section \ref{history}, we consider the concept {\it The Animal eats the Food} as a combination of the individual concepts {\it Animal} and {\it Food}. Let us then consider two pairs of items of {\it Animal}, namely, ({\it Cat}, {\it Cow}) and ({\it Horse}, {\it Squirrel}), and two pairs of items of {\it Food}, namely, ({\it Grass}, {\it Meat}) and ({\it Fish}, {\it Nuts}). The test can be illustrated as follows.

A sample of 81 individuals were presented in a `within subjects design' a questionnaire which contained again the 4 coincidence experiments $AB$, $AB'$, $A'B$ and $A'B'$ that are typical of a Bell-type test. More precisely, participants, after reading an introductory text on concepts and their combinations as in the first test, were asked to choose in each coincidence experiment one item in a list of 4 items, namely, the item that they judged as a good example of the conceptual combination  {\it The Animal eats the Food}.

In coincidence experiment $AB$, participants had to choose among the 4 items: 

($A_1B_1$) {\it The Cat eats the Grass} 

($A_2B_2$) {\it The Cat eats the Meat}

($A_1B_2$) {\it The Cow eats the Grass} 

($A_2B_1$) {\it The Cow eats the Meat}

If the response was ($A_1B_1$) or ($A_2B_2$), then experiment $AB$ was attributed outcome $+1$; if the response was ($A_1B_2$) or ($A_2B_1$), then experiment $AB$ was attributed outcome $-1$. 

In coincidence experiment $AB'$, participants had to choose among the 4 items: 

($A_1B'_1$) {\it The Cat eats the Fish}

($A_1B'_2$) {\it The Cat eats the Nuts} 

($A_2B'_1$) {\it The Cow eats the Fish} 

($A_2B'_2$) {\it The Cow eats the Nuts}

If the response was ($A_1B'_1$) or ($A_2B'_1$), then experiment $AB'$ was attributed outcome $+1$; if the response was ($A_1B'_2$) or ($A_2B'_1$), then experiment $AB'$ was attributed outcome $-1$. 

In coincidence experiment $A'B$, participants had to choose among the 4 items: 

($A'_1B_1$) {\it The Horse eats the Grass}

($A'_1B_2$) {\it The Squirrel eats the Meat}

($A'_2B_1$) {\it The Horse eats the Grass}

($A'_2B_2$) {\it The Cow eats the Meat}

If the response was ($A'_1B_1$) or ($A'_2B_2$), then experiment $A'B$ was attributed outcome $+1$; if the response was ($A'_1B_2$) or ($A'_2B_1$), then experiment $A'B$ was attributed outcome $-1$. 

Finally, in coincidence experiment $A'B'$, participants had to choose among the 4 items: 

($A'_1B'_1$) {\it The Horse eats the Fish}
 
($A'_1B'_2$) {\it The Horse eats the Nuts}

($A'_2B'_1$) {\it The Squirrel eats the Fish}

($A'_2B'_2$) {\it The Squirrel eats the Nuts}

If the response was ($A'_1B'_1$) or ($A'_2B'_2$), then experiment $A'B'$ was attributed outcome $+1$; if the response was ($A'_1B'_2$) or ($A'_2B'_1$), then experiment $A'B'$ was attributed outcome $-1$. 

As in the first test, after collecting relative frequencies of responses, we calculated in the large number limit the probability $\mu(X_iY_j)$ that the outcome $X_iY_j$ is obtained in the coincidence experiment $XY$, $i,j=1,2$, $X=A,A'$, $Y=B,B'$. Table \ref{tab2} reports the judgement probabilities computed in this way. The corresponding expectation values, or correlation functions, are then calculated as follows:
\begin{eqnarray}
E(AB)&=&\mu(A_1B_1)-\mu(A_1B_2)-\mu(A_2B_1)+\mu(A_2B_2)=-0.7284 \label{AB_Test_2} \\
E(AB')&=&\mu(A_1B'_1)-\mu(A_1B'_2)-\mu(A_2B'_1)+\mu(A_2B'_2)=0.8025 \label{AB'_Test_2}  \\
E(A'B)&=&\mu(A'_1B_1)-\mu(A'_1B_2)-\mu(A'_2B_1)+\mu(A'_2B_2)=0.9012 \label{A'B_Test_2} \\
E(A'B')&=&\mu(A'_1B'_1)-\mu(A'_1B'_2)-\mu(A'_2B'_1)+\mu(A'_2B'_2)=0.8519 \label{A'B'_Test_2}
\end{eqnarray}
Inserting Equations (\ref{AB_Test_2})--(\ref{A'B'_Test_2}) into Equation (\ref{chsh}), we get $\Delta_{CHSH}=3.2840 \approx 3.28$. 
\begin{table} \label{tab2}
\centering
\begin{tabular}{|c |c | c | c| c| }
\hline
Experiment $AB$ & \emph{Cat eats Grass} & \emph{Cat eats Meat} & \emph{Cow eats Fish} & \emph{Cow eats Meat}\\
Probability & $\mu(A_1B_1)=0.1111$ & $\mu(A_1B_2)=0.3457$ & $\mu(A_2B_1)=0.5185$ & $\mu(A_2B_2)=0.0247$  \\
\hline
 Experiment $AB'$  & \emph{Cat eats Fish} & \emph{Cat eats Nuts} & \emph{Cow eats Fish} & \emph{Cow eats Nuts}\\
Probability & $\mu(A_1B'_1)=0.8765$ & $\mu(A_1B'_2)=0.0494$ & $\mu(A_2B'_1)=0.0494$   & $\mu(A_2B'_2)=0.0247$ \\
\hline
Experiment $A'B$  &  \emph{Horse eats Grass} & \emph{Horse eats Meat} & \emph{Squirrel eats Grass} & \emph{Squirrel eats Meat}\\
Probability &   $\mu(A'_1B_1)=0.8889$ & $\mu(A'_1B_2)=0.0494$ & $\mu(A'_2B_1)=0$  &  $\mu(A'_2B_2)=0.0617$ \\
\hline
Experiment $A'B'$  &   \emph{Horse eats Fish} & \emph{Horse eats Nuts} & \emph{Squirrel eats Fish} & \emph{Squirrel eats Nuts}\\
Probability &  $\mu(A'_1B'_1)=0.0494$ & $\mu(A'_1B'_2)=0.1235$ & $\mu(A'_2B'_1)=0.0617$ & $\mu(A'_2B'_2)=0.8765$\\
\hline
\end{tabular}
\caption{The data collected in coincidence experiments on entanglement in the conceptual combination {\it The Animal eats the Food}.}
\end{table}

The numerical value 3.28 violates both the CHSH inequality and the Cirel'son bound. In both cases, the deviation is statistically significant (p-values $2.81*10^{-11}$ and $4.40*10^{-2}$, respectively). Hence, this deviation confirms and strengthens the empirical patterns identified in document retrieval tests on the web in \cite{aerts2010} and \cite{beltrangeriente2018}. Also in this case, entanglement between the concepts {\it Animal} and {\it Food} is a natural candidate to explain this non-classical behaviour, due to the meaning connections existing between {\it Animal} and {\it Food}.

We will show in the next two sections that these empirical findings can be represented in the Hilbert space formalism of quantum theory, independently of their behaviour with respect to the  Cirel'son bound. However, we firstly need to present the essentials of a general quantum-theoretic framework that we have recently elaborated to model Bell-type situations in any empirical domain.

\section{Construction of a general Hilbert space model\label{model}}
In this section we study how a theoretical framework can be constructed which models the two cognitive situations in Section \ref{data} but is also general enough to cope with any domain in which Bell-type tests are performed. We will show that the violations of the CHSH inequality and the Cirel'son bound can be simultaneously explained by assuming that a strong form of `quantum entanglement' is present in conceptual combinations. We refine and generalise the modelling scheme that we have elaborated in \cite{aertssozzo2014a,aertssozzo2014b} to identify the entanglement that occurs in Bell-type settings. The theoretical scheme has already been applied with success to model the web tests in \cite{aertsbeltrangerientesozzo2019} (see also Section \ref{history}). We refer to \cite{aertsetal2019} for a detailed analysis and comparison of the entanglement situations occurring in physical and cognitive realms.

The general modelling scheme for Bell-type situations consists in the implementation of three main steps which we summarise in the following.

(i) One identifies in the empirical situation under investigation the composite conceptual entity and the individual conceptual entities composing it.

(ii) One recognises in the composite conceptual entity the states, measurements and outcome probabilities that are relevant to the phenomenon under study.

(iii) One represents entities, states, measurements and outcome probabilities using the Hilbert space representation of entities, states, measurements and outcome probabilities of quantum theory. 

The application of steps (i)--(iii) to {\it The Animal Acts} situation has been presented in various papers \cite{aertssozzo2011,aertssozzo2014a,aertsbeltrangerientesozzo2019}. Let us then focus here on how (i)--(iii) are applied to {\it The Animal eats the Food} situation. Let us start by coincidence experiment $AB$ and its outcomes {\it The Cat eats the Grass}, {\it The The Cat eats the Meat}, {\it The Cow eats the Grass} and {\it The Cow eats the Meat}. 

(i) The conceptual combination {\it The Animal eats the Food} can be considered as a composite conceptual entity made up of the individual conceptual entities {\it Animal} and {\it Food}. 

(ii) Whenever an individual reads the introductory text which explains the details of the test and nature of the involved concepts, this set of instructions prepares the composite entity {\it The Animal eats the Food} in an initial state $p$ which describes the general situation of an animal that eats food. This initial state is the unique conceptual state which all participants are confronted with in the test. In coincidence experiment $AB$, each respondent interacts with this uniquely prepared state $p$ and operates as a measurement context $e_{AB}$  for {\it The Animal eats the Food} which changes $p$ into a generally different state. The latter state is not predetermined, as it depends on the concrete choice being made, which results as a consequence of this `contextual interaction' between the respondent and the conceptual entity. More specifically, if the respondent chooses {\it The Cow eats the Grass}, that is, the outcome $A_2B_1$ is obtained in $AB$ (see Section \ref{data}), the interaction between the entity  {\it The Animal eats the Food} prepared in the state $p$ and the (mind of the) respondent will determine a change of state of {\it The Animal eats the Food} from $p$ to the state $p_{A_2B_1}$ which describes the more concrete situation of a cow that eats grass. When all responses are collected, a statistics of outcomes arises from this intrinsically and genuinely indeterministic contextual process of state change. This outcome statistics is interpreted, in the large number limit, as the probability $P_{p}(A_2B_1)$ that the outcome $A_2B_1$ is obtained when the measurement $e_{AB}$ is performed on the composite entity {\it The Animal eats the Food} in the initial state $p$.\footnote{In \cite{aertssassolidebianchisozzo2016}),  we have called `realistic' and `operational' the above description of a measurement process on a conceptual entity, where the term `realistic' refers to the preparation of the conceptual entity in a defined state and the term `operational' refers to the measurements that are performed on the entity.} 

(iii) Let us finally come to the quantum mathematical representation. The entity {\it The Animal eats the Food} is associated with a Hilbert space and the state $p$ is represented by a unit vector $|p\rangle$ of this Hilbert space. The measurement $e_{AB}$ is instead represented by a self-adjoint operator or, equivalently, by a spectral family, on the Hilbert space whose eigenvectors represent the outcome states, or `eigenstates', of $e_{AB}$, while outcome probabilities are obtained through the Born rule of quantum probability. This representation will become clear in Section \ref{quantumrep}. 

The description above can be extended in a straightforward way to the other coincidence experiments of {\it The Animal eats the Food} situation. Hence, the complete identification of states, measurements, outcomes and outcome probabilities that are relevant to the situations in Section \ref{data} is presented in the following.

For every $X=A,A'$, $Y=B,B'$, coincidence experiment $XY$ corresponds to a measurement $e_{XY}$ performed on the composite conceptual entity {\it The Animal Acts} (respectively, {\it The Animal eats the Food}) with 4 possible outcomes $X_iY_j$, $i,j=1,2$, where we choose $X_iY_j=+1$ if $i=j$ and $X_iY_j=-1$ if $i \ne j$, and 4 eigenstates $p_{X_iY_j}$ describing the state of {\it The Animal Acts} ({\it The Animal eats the Food})  after the outcome $X_iY_j$ occurs in $XY$. Let us denote by $P_{p}(X_iY_j)$ the probability that the outcome $X_iY_j$ is obtained when the measurement $e_{XY}$ is performed on {\it The Animal Acts} ({\it The Animal eats the Food}) in the state $p$.

Once we have completed the identification of entities, initial state, measurements, eigenstates and outcome probabilities occurring in the {\it The Animal Acts} and {\it The Animal eats the Food} situations, we need to construct a general quantum framework in Hilbert space to model these conceptual situations. However, before doing so, we need to recall a theoretical analysis that we have presented in detail in \cite{aertsetal2018e} (see also \cite{aertssozzo2014a,aertsbeltrangerientesozzo2019}).

We preliminarily observe that in both  {\it The Animal Acts}  and {\it The Animal eats the Food} situations, all measurements $e_{XY}$, $X=A,A'$, $Y=B,B'$, have 4 outcomes $X_iY_j$, $i,j=1,2$, which entails that both composite entities should be associated, as overall entities, with the complex Hilbert space $\mathbb{C}^{4}$ of all ordered 4-tuples of complex numbers. In addition, each state $p$ of {\it The Animal Acts}  (respectively, {\it The Animal eats the Food}) should be represented by a unit vector of $\mathbb{C}^{4}$ and each measurement on {\it The Animal Acts}  ({\it The Animal eats the Food}) should be represented by a self-adjoint operator or, equivalently, by a spectral family, on $\mathbb{C}^{4}$. 
On the other side, for every $i,j=1,2$, each outcome $X_iY_j$ is obtained by juxtaposing the outcomes $X_i$ and $Y_j$, e.g., {\it The Bear Growls} is obtained by syntactically juxtaposing the words ``bear'' and ``growls''. This operation defines a 2-outcome measurement $e_{X}$, $X=A,A'$ on the individual entity {\it Animal} ({\it Animal}) and a 2-outcome measurement $e_{Y}$, $Y=B,B'$ on the individual entity {\it Acts} ({\it Food}). Hence, each of these individual entities should be associated with the complex Hilbert space $\mathbb{C}^{2}$ of all ordered couples of complex numbers. 
But, then, the standard Hilbert space formalism prescribes that both composite entities {\it The Animal Acts}  and {\it The Animal eats the Food} should be associated with the tensor product Hilbert space $\mathbb{C}^{2} \otimes \mathbb{C}^{2}$. We stress that we are studying here an `identification problem', that is, the problem of how the composite entity {\it The Animal Acts} ({\it The Animal eats the Food}) can be decomposed into the individual entities {\it Animal} ({\it Animal}) and {\it Acts} ({\it Food})' in such a way that `these  individual entities can be recognised from measurements performed on the respective composite entities'. As such, we are doing an operation that is opposite to what one typically does in Bell-type situations in quantum physics, where one constructs or, better, `composes', the measurements on the composite entity from measurements performed on individual entities. 

From a mathematical point of view, the vector spaces $\mathbb{C}^{4}$ and $\mathbb{C}^{2} \otimes \mathbb{C}^{2}$ are isomorphic, and any isomorphism can be expressed in terms of the relationship between the corresponding orthonormal (ON) bases. States of {\it The Animal Acts} are represented by unit vectors of $\mathbb{C}^{4}$, hence of $\mathbb{C}^{2} \otimes \mathbb{C}^{2}$, which contains both vectors representing `product states' and vectors representing `entangled states'. Moreover, the vector space $L(\mathbb{C}^{4})$ of all linear operators on $\mathbb{C}^{4}$ is isomorphic to the tensor product $L(\mathbb{C}^{2}) \otimes L(\mathbb{C}^{2})$, where $L(\mathbb{C}^{2})$ of all linear operators on $\mathbb{C}^{2}$. Analogously, the tensor product  $L(\mathbb{C}^{2}) \otimes L(\mathbb{C}^{2})$ contains both self-adjoint operators representing `product measurements' and self-adjoint operators representing `entangled measurements'.

Now, let $I: \mathbb{C}^{4} \longrightarrow \mathbb{C}^{2} \otimes \mathbb{C}^{2}$ be an isomorphism mapping a given ON basis of $\mathbb{C}^{4}$ onto a given ON basis of $\mathbb{C}^{2} \otimes \mathbb{C}^{2}$. We say that a state $p$ represented by the unit vector $|p\rangle \in {\mathbb C}^4$ is a `product state' with respect to $I$, if two states $p_A$ and $p_B$, represented by the unit vectors $|p_A\rangle \in {\mathbb C}^2$ and $|p_B\rangle \in {\mathbb C}^2$, respectively, exist such that $I|p\rangle=|p_A\rangle\otimes|p_B\rangle$. Otherwise, $p$ is an `entangled state' with respect to $I$. Next, we say that a measurement $e$ represented by the self-adjoint operator ${\mathscr E}$ on ${\mathbb C}^4$ is a `product measurement' with respect to $I$, if two measurements $e_X$ and $e_Y$, represented by the self-adjoint operators  ${\mathscr E}_X$ and ${\mathscr E}_Y$, respectively, on ${\mathbb C}^2$ exist such that $I{\mathscr E}I^{-1}={\mathscr E}_X \otimes {\mathscr E}_Y$. Otherwise, $e$ is an `entangled measurement' with respect to $I$. Hence, the notion of entanglement crucially depends on the `isomorphism that is used to identify individual entities within a given composite entity'. 

With reference to the Bell-type setting presented in this section, one can then prove that, if the measurements $e_{XY}$ and  $e_{XY'}$, $X=A,A'$, $Y,Y'=B,B'$, $Y' \ne Y$, are product measurements with respect to the isomorphism $I$, then, for every state $p$ of the composed entity,  the `marginal law of Kolmogorovian probability' is satisfied, that is, for every $i=1,2$, $\sum_{j} P_{p}(X_iY_j)=\sum_{j}P_{p}(X_iY'_j)$. Analogously, if the measurements $e_{XY}$ and  $e_{X'Y}$, $X,X'=A,A'$, $Y=B,B'$, $X' \ne X$, are product measurements with respect to the isomorphism $I$, then,  for every state $p$ of the composed entity,  the marginal law is satisfied, that is, for every $j=1,2$, $\sum_{i} P_{p}(X_iY_j)=\sum_{i}P_{p}(X'_iY_j)$ (see Theorem 2, \cite{aertssozzo2014a}). In the case the marginal law is satisfied in all measurements, one can also prove that a unique isomorphism exists, which can be chosen to be the identify operator (see, Theorem 4, \cite{aertssozzo2014a}).

It follows from the above that, if the marginal law is violated, as it occurs in both conceptual tests in Section \ref{data},\footnote{The marginal law is systematically violated in {\it The Animal Acts} test. For example, $\mu(A_2B_1)+\mu(A_2B_2)=0.8272\ne 0.2345=\mu(A_2B'_1)+\mu(A_2B'_2)$. Analogously, the marginal law is systematically violated in {\it The Animal eats the Food} test. For example, $\mu(A'_1B_1)+\mu(A'_1B_2)=0.9383\ne 0.1729=\mu(A'_1B'_1)+\mu(A'_1B'_2)$.} then one cannot find a unique isomorphism between $\mathbb{C}^{4}$ and $\mathbb{C}^{2} \otimes \mathbb{C}^{2}$ such that all measurements are product measurements with respect to this isomorphism. In this case, one cannot explain the violation of the CHSH inequality as due to the usual situation in quantum physics where all measurements are product measurements and only the initial, or pre-measurement, state is entangled. Furthermore, if the marginal law is systematically violated, 4 distinct isomorphisms $I_{XY}$, $X=A,A'$, $Y=B,B'$, exist such that the measurement  $e_{XY}$ is a product measurement with respect to $I_{XY}$. As a consequence, there is no unique isomorphism allowing to identify individual entities of a given composite entity. Finally, if we consider a given isomorphism between $\mathbb{C}^{4}$ and $\mathbb{C}^{2} \otimes \mathbb{C}^{2}$ with respect to which identifying individual entities of a composite entity in a given test, then it may happen that both the pre-measurement state and all measurements are entangled. This final remark suggests the following considerations.

Firstly, the non-classical connections which violate the CHSH inequality in both {\it The Animal Acts} and {\it The Animal eats the Food} situations can be reasonably attributed to the fact that `the component individual concepts carry meaning and further meaning is created in the combination process'. Since the violation of the CHSH inequality indicates the presence of entanglement between the individual conceptual entities, then it is reasonable to assume that `it is the quantum structure of entanglement which is able to theoretically capture the meaning connections that are created in these cases'. This suggests, in a quantum-theoretic perspective, that the initial state $p$ of both composite entities  {\it The Animal Acts} and {\it The Animal eats the Food} should be an entangled state. This entangled state would then capture the meaning connections that are created between {\it Animal} and {\it Acts} and also between {\it Animal} and {\it Food} when the respondent read the introductory text and start the questionnaire.

Secondly, in both {\it The Animal Acts} and {\it The Animal eats the Food} situations, all measurements $e_{XY}$, $X=A,A'$, $Y=B,B'$, violate the marginal law of Kolmogorovian probability. This suggests, again in a quantum-theoretic perspective, that all these measurements should be entangled measurements. In addition, in each measurement, all outcomes $X_iY_j$, $i,j=1,2$, correspond to combined concepts, e.g., {\it The Cat Meows}, {\it The Cat eats the Fish}, etc., which are thus in turn connected by meaning. This also suggests that all eigenstates $p_{X_iY_j}$ should be entangled states.

The two considerations above applied to the Hilbert space representation of both {\it The Animal Acts} cognitive and web tests \cite{aertssozzo2014a,aertsbeltrangerientesozzo2019}. We will see in the next section that these considerations also apply to the empirical tests in  in Section \ref{data}.

\section{Quantum representation of data\label{quantumrep}}
We work out in this section a quantum representation in Hilbert space of the cognitive tests data in Section \ref{data}, following the methodology developed in \cite{aertsetal2019} and already applied to previous tests \cite{aertssozzo2014a,aertsbeltrangerientesozzo2019}. The considerations in Section \ref{model} suggest the following quantum mathematical representation for both {\it The Animal Acts} and {\it The Animal eats the Food} situations.

The composite conceptual entity is associated with the Hilbert space $\mathbb{C}^4$ of all ordered 4-tuples of complex numbers. Let 
$(1,0,0,0)$, $(0,1,0,0)$, $(0,0,1,0)$ and $(0,0,0,1)\}$ be the unit vectors of the canonical ON basis of $\mathbb{C}^4$ and let us consider the isomorphism $I:\mathbb{C}^4  \longrightarrow \mathbb{C}^2 \otimes \mathbb{C}^2$ where this ON basis coincides with the ON basis of the tensor product Hilbert space $\mathbb{C}^2 \otimes \mathbb{C}^2$ made up of the unit vectors $(1,0)\otimes (1,0)$, $(1,0)\otimes (0,1)$, $(0,1)\otimes (1,0)$ and $(0,1)\otimes (0,1)$. In these ON bases, the initial state $p$ of the composite entity is represented by the unit vector $|p\rangle=(ae^{i \alpha}, be^{i \beta}, ce^{i \gamma}, de^{i \delta})$, where $a,b,c,d \ge 0$, $a^2+b^2+c^2+d^2=1$, $\alpha$, $\beta$, $\gamma$, $\delta \in \Re$ and $\Re$ is the real line. One easily proves that $|p\rangle$ represents a product state if and only if the following condition is satisfied:
\begin{equation} \label{entanglementcondition}
ade^{i(\alpha+\delta)}-bce^{i(\beta+\gamma)}=0
\end{equation}
Otherwise, $p$ represents an entangled state.

Let us now represent the measurements $e_{AB}$, $e_{AB'}$, $e_{A'B}$ and $e_{A'B'}$ in Section \ref{model}. Each measurement $e_{XY}$, $X=A,A'$, $Y=B,B'$, has 4 outcomes $X_1Y_1$, $X_1Y_2$, $X_2Y_1$ and $X_2Y_2$ and 4 eigenstates $p_{X_1Y_1}$, $p_{X_1Y_2}$, $p_{X_2Y_1}$ and $p_{X_2Y_2}$. As in Sections \ref{data} and \ref{model}, we set, for every $X=A,A'$, $Y=B,B'$, $X_1Y_{1}=X_2Y_{2}=+1$ and $X_1Y_{2}=X_2Y_{1}=-1$.
 The measurement  $e_{XY}$ is represented by the self-adjoint operator ${\mathscr E}_{XY}$ on ${\mathbb C}^4$ or, equivalently, by the spectral family $\{|p_{X_1Y_1}\rangle\langle p_{X_1Y_1}|, |p_{X_1Y_2}\rangle\langle p_{X_1Y_2}|, |p_{X_2Y_1}\rangle\langle p_{X_2Y_1}|, |p_{X_2Y_2}\rangle\langle p_{X_2Y_2}| \}$, such that the eigenstates $p_{X_1Y_1}$, $p_{X_1Y_2}$, $p_{X_2Y_1}$ and $p_{X_2Y_2}$ are represented by the eigenvectors
\begin{eqnarray}
|p_{11}\rangle&=&(a_{11}e^{i \alpha_{11}}, b_{11}e^{i \beta_{11}}, c_{11}e^{i \gamma_{11}}, d_{11}e^{i \delta_{11}}) \label{HG}\\
|p_{12}\rangle&=&(a_{12}e^{i \alpha_{12}}, b_{12}e^{i \beta_{12}}, c_{12}e^{i \gamma_{12}}, d_{12}e^{i \delta_{12}}) \label{HW}\\
|p_{21}\rangle&=&(a_{21}e^{i \alpha_{21}}, b_{21}e^{i \beta_{21}}, c_{21}e^{i \gamma_{21}}, d_{21}e^{i \delta_{21}}) \label{BG}\\
|p_{22}\rangle&=&(a_{22}e^{i \alpha_{22}}, b_{22}e^{i \beta_{22}}, c_{22}e^{i \gamma_{22}}, d_{22}e^{i \delta_{22}}) \label{BW}
\end{eqnarray}
of ${\mathscr E}_{XY}$, respectively. In Equations (\ref{HG})--(\ref{BW}), the coefficients are such that $a_{ij}, b_{ij}, c_{ij}, d_{ij} \ge 0$ and $\alpha_{ij}, \beta_{ij},\gamma_{ij}, \delta_{ij} \in\Re$, $i,j=1,2$. For every $X=A,A'$, $Y=B,B'$, the self-adjoint operator ${\mathscr E}_{XY}$ can be expressed as a tensor product operator if and only if all unit vectors in Equations (\ref{HG})--(\ref{BW}) represent product states. Otherwise, ${\mathscr E}_{XY}$ cannot be expressed as a tensor product operator, hence $e_{XY}$ is an entangled measurement. For every $X=A,A'$, $Y=B,B'$, $i,j=1,2$, the probability $P_{p}(X_iY_j)$ of obtaining the outcome $X_iY_j$ in a measurement of $e_{XY}$ on the composite entity in the state $p$ is then given by the Born rule of quantum probability, that is, $P_{p}(X_iY_j)=|\langle p_{ij}|p\rangle|^2$.
 
To find a quantum mathematical representation of the data in Section \ref{data}, for every measurement $e_{XY}$, the unit vectors in (\ref{HG})--(\ref{BW}) have to satisfy the following three sets of conditions.

(i) Normalization. The vectors in Equations (\ref{HG})--(\ref{BW})  are unitary, that is,
\begin{eqnarray}
a_{11}^2+b_{11}^2+c_{11}^2+d_{11}^2&=&1  \\
a_{12}^2+b_{12}^2+c_{12}^2+d_{12}^2&=&1 \\
a_{21}^2+b_{21}^2+c_{21}^2+d_{21}^2&=&1  \\
a_{22}^2+b_{22}^2+c_{22}^2+d_{22}^2&=&1
\end{eqnarray}

(ii) Orthogonality. The vectors in Equations (\ref{HG})--(\ref{BW})  are mutually orthogonal, that is,
\begin{eqnarray}
0=\langle p_{11}|p_{12} \rangle=a_{11}a_{12}e^{i(\alpha_{12}-\alpha_{11})}+b_{11}b_{12}e^{i(\beta_{12}-\beta_{11})}+a_{11}c_{12}c^{i(\gamma_{12}-\gamma_{11})}+d_{11}d_{12}e^{i(\delta_{12}-\delta_{11})}  \\
0=\langle p_{11}|p_{21} \rangle=a_{11}a_{21}e^{i(\alpha_{21}-\alpha_{11})}+b_{11}b_{21}e^{i(\beta_{21}-\beta_{11})}+a_{11}c_{21}c^{i(\gamma_{21}-\gamma_{11})}+d_{11}d_{21}e^{i(\delta_{21}-\delta_{11})}  \\
0=\langle p_{11}|p_{22} \rangle=a_{11}a_{22}e^{i(\alpha_{22}-\alpha_{11})}+b_{11}b_{22}e^{i(\beta_{22}-\beta_{11})}+a_{11}c_{22}c^{i(\gamma_{22}-\gamma_{11})}+d_{11}d_{22}e^{i(\delta_{22}-\delta_{11})}  \\
0=\langle p_{12}|p_{21} \rangle=a_{12}a_{21}e^{i(\alpha_{21}-\alpha_{12})}+b_{12}b_{21}e^{i(\beta_{21}-\beta_{12})}+a_{12}c_{21}c^{i(\gamma_{21}-\gamma_{12})}+d_{12}d_{21}e^{i(\delta_{21}-\delta_{12})}  \\
0=\langle p_{12}|p_{22} \rangle=a_{12}a_{22}e^{i(\alpha_{22}-\alpha_{12})}+b_{12}b_{22}e^{i(\beta_{22}-\beta_{12})}+a_{12}c_{22}c^{i(\gamma_{22}-\gamma_{12})}+d_{12}d_{22}e^{i(\delta_{22}-\delta_{12})}  \\
0=\langle p_{21}|p_{22} \rangle=a_{21}a_{22}e^{i(\alpha_{22}-\alpha_{21})}+b_{21}b_{22}e^{i(\beta_{22}-\beta_{21})}+a_{21}c_{22}c^{i(\gamma_{22}-\gamma_{21})}+d_{21}d_{22}e^{i(\delta_{22}-\delta_{21})}
\end{eqnarray}

(iii) Probabilities. For every $X=A,A'$, $Y=B,B'$, $i,j=1,2$, the probability $P_{p}(X_iY_j)$ coincides with the empirical probability $\mu(X_iY_j)$ in Tables 1 and 2, that is,
\begin{eqnarray}
\mu(X_1Y_1)=|\langle p_{11}|p\rangle|^{2}&=&a^2 a_{11}^2+b^2 b_{11}^2+c^2 c_{11}^2+d^2 d_{11}^2+ \nonumber \\
&+&2aba_{11}b_{11}\cos(\alpha-\alpha_{11}-\beta+\beta_{11})+ \nonumber \\
&+&2aca_{11}c_{11}\cos(\alpha-\alpha_{11}-\gamma+\gamma_{11})+ \nonumber \\
&+&2ada_{11}d_{11}\cos(\alpha-\alpha_{11}-\delta+\delta_{11})+ \nonumber \\
&+&2bcb_{11}c_{11}\cos(\beta-\beta_{11}-\gamma+\gamma_{11})+ \nonumber \\
&+&2bdb_{11}d_{11}\cos(\beta-\beta_{11}-\delta+\delta_{11})+ \nonumber \\
&+&2cdc_{11}d_{11}\cos(\gamma-\gamma_{11}-\delta+\delta_{11})
\end{eqnarray}
\begin{eqnarray}
\mu(X_1Y_2)=|\langle p_{12}|p\rangle|^{2}&=&a^2 a_{12}^2+b^2 b_{12}^2+c^2 c_{12}^2+d^2 d_{12}^2+ \nonumber \\
&+&2aba_{12}b_{12}\cos(\alpha-\alpha_{12}-\beta+\beta_{12})+ \nonumber \\
&+&2aca_{12}c_{12}\cos(\alpha-\alpha_{12}-\gamma+\gamma_{12})+ \nonumber \\
&+&2ada_{12}d_{12}\cos(\alpha-\alpha_{12}-\delta+\delta_{12})+ \nonumber \\
&+&2bcb_{12}c_{12}\cos(\beta-\beta_{12}-\gamma+\gamma_{12})+ \nonumber \\
&+&2bdb_{12}d_{12}\cos(\beta-\beta_{12}-\delta+\delta_{12})+ \nonumber \\
&+&2cdc_{12}d_{12}\cos(\gamma-\gamma_{12}-\delta+\delta_{12}) 
\end{eqnarray}
\begin{eqnarray}
\mu(X_2Y_1)=|\langle p_{21}|p\rangle|^{2}&=&a^2 a_{21}^2+b^2 b_{21}^2+c^2 c_{21}^2+d^2 d_{21}^2+ \nonumber \\
&+&2aba_{21}b_{21}\cos(\alpha-\alpha_{21}-\beta+\beta_{21})+ \nonumber \\
&+&2aca_{21}c_{21}\cos(\alpha-\alpha_{21}-\gamma+\gamma_{21})+ \nonumber \\
&+&2ada_{21}d_{21}\cos(\alpha-\alpha_{21}-\delta+\delta_{21})+ \nonumber \\
&+&2bcb_{21}c_{21}\cos(\beta-\beta_{21}-\gamma+\gamma_{21})+ \nonumber \\
&+&2bdb_{21}d_{21}\cos(\beta-\beta_{21}-\delta+\delta_{21})+ \nonumber \\
&+&2cdc_{21}d_{21}\cos(\gamma-\gamma_{21}-\delta+\delta_{21})
\end{eqnarray}
\begin{eqnarray}
\mu(X_2Y_2)=|\langle p_{22}|p\rangle|^{2}&=&a^2 a_{22}^2+b^2 b_{22}^2+c^2 c_{22}^2+d^2 d_{22}^2+ \nonumber \\
&+&2aba_{22}b_{22}\cos(\alpha-\alpha_{22}-\beta+\beta_{22})+ \nonumber \\
&+&2aca_{22}c_{22}\cos(\alpha-\alpha_{22}-\gamma+\gamma_{22})+ \nonumber \\
&+&2ada_{22}d_{22}\cos(\alpha-\alpha_{22}-\delta+\delta_{22})+ \nonumber \\
&+&2bcb_{22}c_{22}\cos(\beta-\beta_{22}-\gamma+\gamma_{22})+ \nonumber \\
&+&2bdb_{22}d_{22}\cos(\beta-\beta_{22}-\delta+\delta_{22})+ \nonumber \\
&+&2cdc_{22}d_{22}\cos(\gamma-\gamma_{22}-\delta+\delta_{22})
\end{eqnarray}

Then, let us represent represent the initial state $p$ of the composite conceptual entity by the unit vector
\begin{equation}
|p\rangle=\frac{1}{\sqrt{2}}(0,1,-1,0) \label{singlet}
\end{equation}
The vector in Equation (\ref{singlet}) represents a maximally entangled state and corresponds to the `singlet spin state' that is used in typical Bell-type tests in quantum physics (see Section \ref{intro}). There are different reasons for this choice. Firstly, our aim is to incorporate all possible entanglement of the state-measurement situation in the pre-measurement state, so that the ensuing measurements are as close as possible to product measurements. We are however aware that some measurements, if not all, are entangled, due to the violation of the marginal law (see Section \ref{model}). Secondly, it is known that the singlet spin state has specific symmetry properties, namely, it is always represented by a unit vector of the form in Equation (\ref{singlet}) independently of the ON basis in which the unit vector is expressed. This would intuitively correspond to the fact both {\it The Animal Acts} and {\it The Animal eats the Food} express more abstract concepts than the corresponding outcomes.

Thus, for every $X=A,A'$, $Y=B,B'$, conditions (i)--(iii) identify 20 equations which should be satisfied by the 32 real variables $a_{ij}, b_{ij}, c_{ij}, d_{ij}, \alpha_{ij}, \beta_{ij}, \gamma_{ij}, \delta_{ij}$, $i,j=1,2$. To simplify the calculation, we set, for every $i,j=1,2$, $\alpha_{ij}=\beta_{ij}=\gamma_{ij}=\delta_{ij}=\theta_{ij}$, where $\theta_{ij}\in \Re$. Thus, each unit vector in Equations (\ref{HG})--(\ref{BW}) takes the form $|p_{ij}\rangle =e^{i \theta_{ij}}(a_{ij}, b_{ij}, c_{ij}, d_{ij})$, $i,j=1,2$. This reduces the total number of unknown variables to 20.

We are now ready to represent the empirical data in Section \ref{data} in Hilbert space. We start by {\it The Animal Acts} test.

The eigenstates of the measurement $e_{AB}$ are represented by the unit vectors
\begin{eqnarray}
|p_{A_1B_1}\rangle &=&e^{i 174.91^{\circ}}(0.97,-0.10,0.21,0) \label{Gsol_HG} \\
|p_{A_1B_2}\rangle &=&e^{i 0.38^{\circ}}(0,0.88,0.38,0.29) \label{Gsol_HW} \\
|p_{A_2B_1}\rangle &=&e^{i 32.54^{\circ}}(0.23,0.35,-0.90,0.12) \label{Gsol_BG} \\
|p_{A_2B_2}\rangle &=&e^{i 98.90^{\circ}}(0.03,0.32,0,-0.95) \label{Gsol_BW}
\end{eqnarray}
By applying the entanglement condition in Equation (\ref{entanglementcondition}), we can verify that all unit vectors are entangled, hence $e_{AB}$ is an entangled measurement. However, one observes that the condition in Equation (\ref{entanglementcondition}) shows a relatively larger deviation from zero in the unit vector $|p_{A_2B_1}\rangle$. We can then say that the eigenstate $p_{A_2B_1}$, corresponding to {\it The Bear Growls}, is a `relatively more entangled state'.

The eigenstates of the measurement $e_{AB'}$ are represented by the unit vectors
\begin{eqnarray}
|p_{A_1B'_1}\rangle &=&e^{i 0.07^{\circ}}(0.14,0.24,-0.96,0.08)  \label{Gsol_HS} \\
|p_{A_1B'_2}\rangle &=&e^{i 18.24^{\circ}}(0.01,0.31,0,-0.95) \label{Gsol_HM} \\
|p_{A_2B'_1}\rangle &=&e^{i 64.68^{\circ}}(-0.02,0.92,0.25,0.30) \label{Gsol_BS} \\
|p_{A_2B'_2}\rangle &=&e^{i 132.68^{\circ}}(0.99,-0.02,0.14,0) \label{Gsol_BM}
\end{eqnarray}
Also in this case, all unit vectors are entangled, hence $e_{AB'}$ is an entangled measurement. The entanglement condition in Equation (\ref{entanglementcondition}) shows a relatively larger deviation from zero in the unit vector $|p_{A_1B'_1}\rangle$. We can then say that the eigenstate $p_{A_1B'_1}$, corresponding to {\it The Horse Snorts}, is a `relatively more entangled state'.

The eigenstates of the measurement $e_{A'B}$ are represented by the unit vectors
\begin{eqnarray}
|p_{A'_1B_1}\rangle &=&e^{i 45.48^{\circ}}(0.25,0.37,-0.88,0.17)\label{Gsol_TG} \\
|p_{A'_1B_2}\rangle &=&e^{i 2.21^{\circ}}(0.01,0.84,0.42,0.35) \label{Gsol_TW} \\
|p_{A'_2B_1}\rangle &=&e^{i 308.44^{\circ}}(0.97,-0.13,0.22,0) \label{Gsol_CG} \\
|p_{A'_2B_2}\rangle &=&e^{i 163.38^{\circ}}(0.05,0.39,0,-0.92) \label{Gsol_CW}
\end{eqnarray}
All unit vectors are entangled, hence $e_{A'B}$ is an entangled measurement. The entanglement condition in Equation (\ref{entanglementcondition}) shows a relatively larger deviation from zero in the unit vector $|p_{A'_1B_1}\rangle$. We can then say that the eigenstate $p_{A'_1B_1}$, corresponding to {\it The Tiger Growls}, is a `relatively more entangled state'.

Finally, the eigenstates of the measurement $e_{A'B'}$ are represented by the unit vectors
\begin{eqnarray}
|p_{A'_1B'_1}\rangle &=&e^{i 353.32^{\circ}}(0.96,-0.18,0.23,0) \label{Gsol_TS} \\
|p_{A'_1B'_2}\rangle &=&e^{i 143.07^{\circ}}(0.07,0.35,0,-0.93)\label{Gsol_TM} \\
|p_{A'_2B'_1}\rangle &=&e^{i 0.58^{\circ}}(0.01,0.78,0.55,0.30) \label{Gsol_CS} \\
|p_{A'_2B'_2}\rangle &=&e^{i 41.95^{\circ}}(0.29,0.49,-0.80,0.20) \label{Gsol_CM}
\end{eqnarray}
All unit vectors are entangled, hence $e_{A'B'}$ is an entangled measurement.  The entanglement condition in Equation (\ref{entanglementcondition}) shows a relatively larger deviation from zero in the unit vector $|p_{A'_2B'_2}\rangle$. We can then say that the eigenstate $p_{A'_2B'_2}$, corresponding to {\it The Cat Meows} , is a `relatively more entangled state'.

Let us now come to the representation of empirical data of {\it The Animal eats the Food} test. The eigenstates of the measurements $e_{AB}$, $e_{AB'}$, $e_{A'B}$ and $e_{A'B'}$ are respectively represented by the 4 unit vectors
\begin{eqnarray}
|p_{A_1B_1}\rangle &=&e^{i 156.99^{\circ}}(0.91,-0.06,0.41,0) \label{Csol_HG} \\
|p_{A_1B_2}\rangle &=&e^{i 0.38^{\circ}}(0.01,0.97,0.13,0.22) \label{Csol_HW} \\
|p_{A_2B_1}\rangle &=&e^{i 117.49^{\circ}}(0.41,0.11,-0.90,0.03)\label{Csol_BG} \\
|p_{A_2B_2}\rangle &=&e^{i 160.39^{\circ}}(0.02,0.22,0,-0.97)\label{Csol_BW}
\end{eqnarray}
\begin{eqnarray}
|p_{A_1B'_1}\rangle &=&e^{i 45.38^{\circ}}(0.22,0.49,-0.84,0.12)\label{Csol_HS} \\
|p_{A_1B'_2}\rangle &=&e^{i 2.20^{\circ}}(0.01,0.83,0.52,0.19) \label{Csol_HM} \\
|p_{A_2B'_1}\rangle &=&e^{i 308.12^{\circ}}(0.97,-0.13,0.18,0) \label{Csol_BS} \\
|p_{A_2B'_2}\rangle &=&e^{i 158.68^{\circ}}(0.03,0.22,0,-0.97) \label{Csol_BM}
\end{eqnarray}
\begin{eqnarray}
|p_{A'_1B_1}\rangle &=&e^{i 45.43^{\circ}}(0,0.47,-0.87,0.17) \label{Csol_TG} \\
|p_{A'_1B_2}\rangle &=&e^{i 2.21^{\circ}}(0.02,0.81,0.50,0.31)\label{Csol_TW} \\
|p_{A'_2B_1}\rangle &=&e^{i 337.01^{\circ}}(0.99,-0.01,-0.01,0) \label{Csol_CG} \\
|p_{A'_2B_2}\rangle &=&e^{i 163.38^{\circ}}(0,0.35,0,-0.94) \label{Csol_CW}
\end{eqnarray}
\begin{eqnarray}
|p_{A'_1B'_1}\rangle &=&e^{i 353.52^{\circ}}(0.97,-0.13,0.19,0) \label{Csol_TS} \\
|p_{A'_1B'_2}\rangle &=&e^{i 143.07^{\circ}}(0.02,0.16,0,-0.99) \label{Csol_TM} \\
|p_{A'_2B'_1}\rangle &=&e^{i 0.58^{\circ}}(0.02,0.85,0.50,0.14) \label{Csol_CS} \\
|p_{A'_2B'_2}\rangle &=&e^{i 42.01^{\circ}}(0.22,0.48,-0.84,0.08) \label{Csol_CM}
\end{eqnarray}
Also in this case, referring to Equation (\ref{entanglementcondition}), one can show that all unit vectors in Equations (\ref{Csol_HG})--(\ref{Csol_CM}) are entangled, hence `all measurements $e_{AB}$, $e_{AB'}$, $e_{A'B}$ and $e_{A'B'}$ are entangled measurements'. In addition, as in {\it The Animal Acts} test, in each measurement, one eigenstate is represented by a unit vectors which exhibits a larger deviation in the entanglement condition in Equation (\ref{entanglementcondition}). These `relatively more entangled states' are $p_{A_2B_1}$, corresponding to {\it The Cow eats the Grass} in $e_{AB}$,  $p_{A_1B'_1}$, corresponding to {\it The Cat eats the Fish} in $e_{AB'}$, $p_{A'_1B_1}$, corresponding to {\it The Horse eats the Grass} in $e_{A'B}$, and $p_{A'_2B'_2}$, corresponding to {\it The Squirrel eats the Nuts} in $e_{A'B'}$.

We have thus completed the quantum mathematical representation of the data on {\it The Animal Acts} and {\it The Animal eats the Food} tests. This representation also suggests additional considerations, as follows.

We firstly observe that all measurements are entangled in the quantum representation in both {\it The Animal Acts} and {\it The Animal eats the Food} situations. This result is due to the violation of the marginal law of probability in both tests which forbids concentrating all the entanglement of the state-measurement situation into the state, as we have seen in Section \ref{model}. Moreover, in each measurement, all eigenstates are entangled. This result confirms with our suggestion in Section \ref{model} that `quantum entanglement theoretically captures the meaning connections between the individual concepts that form the composite conceptual entity, and that these meaning connections are not predetermined, but are created when the test is performed and each respondent interacts with the composite entity'.

Secondly, in both situations, there are some outcome eigenstates which exhibit a relatively higher degree of entanglement than others, which can exactly be explained with the fact that entanglement captures meaning connections, hence higher degrees of entanglement correspond to higher meaning connections. For example, in {\it The Animal Acts} situation, the eigenstate $p_{A_2B_1}$ of the measurement $e_{AB}$, corresponding to {\it The Bear Growls}, is relatively more entangled than the other eigenstates of $e_{AB}$, which can be naturally explained by the fact that the individual items {\it Bear} and {\it Growls}, which are concepts themselves, are relatively more connected by meaning. As a matter of fact, if we look at empirical probabilities (see Table 1, Section \ref{data}), we note that the outcome $A_2B_1$, corresponding to {\it The Bear Growls}, scores a high probability to be judged as a good example of the conceptual combination {\it The Animal Acts}. Vice versa, items like {\it Bear Meows} and {\it Horse Growls} score a low probability and are less connected by meaning. And, indeed, the corresponding eigenstates are relatively less entangled. Analogously, in {\it The Animal eats the Food} situation, the eigenstate $p_{A'_2B'_2}$ of the measurement $e_{A'B'}$, corresponding to {\it The Squirrel eats the Nuts}, is relatively more entangled than the other outcome eigenstates of $e_{A'B'}$, which can be again explained by the fact that the individual items {\it Squirrel} and {\it Nuts}, which are concepts themselves, are relatively more connected by meaning. And, indeed, if we look at empirical probabilities (see Table 2, Section \ref{data}), we note that the outcome $A'_2B'_2$, corresponding to {\it The Squirrel eats the Nuts}, scores a high probability to be judged as a good example of the conceptual combination {\it The Animal eats the Food}. On the contrary,  items like {\it The Cow eats the Meat}, {\it The Cat eats the Nuts} and {\it The Squirrel eats the Grass} score a low probability and are less meaning-connected. And, indeed, the corresponding eigenstates are relatively less entangled.

Thirdly, we observe that the quantum representation of the cognitive tests appears to be more complex and less symmetric than the quantum representation of the information retrieval tests on the web in \cite{beltrangeriente2018} (see \cite{aertsbeltrangerientesozzo2019}), where  all measurements were entangled but each spectral measure contained 
two product eigenstates and two entangled eigenstates. The reason is that many judgement probabilities were zero in \cite{beltrangeriente2018}, due to the fact that human minds are able to create additional meaning connections between concepts than 
corpuses of documents. 

Fourthly, we observe that the violation of Cirel'son bound in {\it The Animal eats the Food} situation can still be explained in terms of quantum entanglement, in contrast to widespread beliefs. Hence, there is quantum entanglement also beyond Cirel'son bound, but this type of entanglement involves both states and measurements. This result is important and we will devote the next section to deepen it.

\section{Quantum entanglement beyond Cirel'son bound\label{bound}}
The empirical results in the cognitive test on {\it The Animal eats the Food} presented in Section \ref{data} show a significant violation of the Cirel'son bound and, as such, they confirm the results in the document retrieval tests on the web in \cite{aerts2010} and \cite{beltrangeriente2018}. As we have noticed throughout the paper, these results contrast with widespread beliefs in quantum physics. It is thus worth to carefully explain their meaning and implications.

The original question behind the determination of the Cirel'son bound was whether there is a fundamental limit to quantum nonlocality, that is, whether the correlations exhibited by two far away physical entities should satisfy any condition in order to represent them in the mathematical formalism of quantum theory -- this problem is also connected with the old problem of exploiting quantum nonlocality to send a faster than light signal between far away entities (see, e.g., \cite{brunner2014}). One then considers a typical Bell-type situation,  as the one presented in Section \ref{intro}, performs a set of 4 measurements on a composite entity, made up of two far away individual entities, by separately performing each time one measurement on one individual entity and one measurement on the other, and deduces that the CHSH factor in Equation (\ref{chsh}) is bound by the value $\Delta_{QMC}=2\sqrt{2}\approx 2.83$ in quantum theory \cite{cirelson1980,cirelson1993}. Hence, the Cirel'son bound obtained in this way is usually considered, especially by quantum physicists, as the limit outside which the correlations exhibited by any two entities cannot be modelled within the Hilbert space formalism of quantum theory \cite{cirelson1980,cirelson1993}. 

On the other side, we have proved, in this and in other papers that it is possible to represented in Hilbert space empirical data collected in Bell-type tests which violate the CHSH inequality by an amount which also exceeds Cirel'son bound (see, e.g., \cite{aertsbeltrangerientesozzo2019,aertsetal2019}). This means that the statistical correlations exhibited in these Bell-type tests, which are anyway non-classical and non-signalling, can be represented within quantum theory.

As we have seen in Sections \ref{model} and \ref{quantumrep}, the fundamental element which makes it possible the above mentioned quantum mathematical representation in Hilbert space is the suggestion that the measurements performed on the composite entity have outcomes that are products of outcomes obtained in measurements separately performed on individual entities, but their eigenstates are generally entangled, that is, they are entangled measurements, hence they cannot generally be decomposed into product measurements. Cirel'son, and the scholars who investigated this bound after him, did not envisage such alternative, hence they implicitly assumed only  product measurements in the derivation of the Cirel'son bound. If one also allows the possibility of using entangled measurements, then one can in principle violate the CHSH inequality in Equation (\ref{chsh}) by any amount within its mathematical limit, i.e. between $-4$ and $+4$ (see also \cite{aertsetal2019}).

The result above is relevant, in our opinion, for various reasons. Firstly, it follows from a careful investigation of the identification problem in which individual entities are identified through measurements that are performed on composite entities. Secondly, it shows that a violation of Bell inequalities which also exceeds the Cirel'son bound can still be explained in terms of quantum entanglement. More important, this `quantum entanglement beyond Cirel'son bound' is deeper and stronger than the entanglement that is generally believed to be responsible of the violation of Bell inequalities in quantum physics., i.e. `entanglement in the state'. Indeed, a violation of the Cirel'son bound, in addition to a violation of Bell inequalities, can be explained by assuming the presence of entanglement in both the state and the measurements, i.e. `state-measurement entanglement'. Thirdly, together with the violation of the marginal law, the violation of the Cirel'son bound may have deep implications on the foundations of quantum physics, as it could shed new light on the problem of faster than light communication and signalling and could explain apparent deviations from quantum predictions in some Bell-type tests in physics. These deviations have been studied putting forward the hypothesis that they would be due to measurement errors \cite{adenierkhrennikov2007,adenierkhrennikov2016}, however in \cite{aerts2014} it is shown that they might well be of a fundamental nature and not be due to measurement errors. If the analysis put forward in \cite{aerts2014} turns out to be true a whole new perspective is needed also on the physical violation of the Bell inequalities where experimentally results a violation of the marginal law \cite{aspect1982a,weihs1998}, indicating that the type of signalling believed to be possible as a consequence of the violation of the marginal law is `not' the one imagined taking place and hence would not, due to its different nature, be able to be faster than light even if the coincidence events are space-like separated events \cite{aerts2014}.

\section*{Acknowledgements}
This work was supported by QUARTZ (Quantum Information Access and Retrieval Theory), the Marie Sk{\l}odowska-Curie Innovative Training Network 721321 of the European Union's Horizon 2020 research and innovation programme.

\end{document}